# Cognitive Effort in Collective Software Design: Methodological Perspectives in Cognitive Ergonomics

Françoise Détienne, Jean-Marie Burkhardt, Willemien Visser

EIFFEL Research group - Cognition and Cooperation in Design, INRIA,
Domaine de Voluceau, Rocquencourt, 78153 Le Chesnay, France
`{Francoise.Detienne, Jean-Marie.Burkhardt, Willemien.Visser}`
`@inria.fr`

**Abstract.** Empirical software engineering is concerned with measuring, or estimating, both the effort put into the software process and the quality of its product. We defend the idea that measuring process effort and product quality and establishing a relation between the two cannot be performed without a model of cognitive and collective activities involved in software design, and without measurement of these activities. This is the object of our field, i.e. Cognitive Ergonomics of design. After a brief presentation of its theoretical and methodological foundations, we will discuss a cognitive approach to design activities and its potential to provide new directions in ESE. Then we will present and discuss an illustration of the methodological directions we have proposed for the analysis and measurement of cognitive activities in the context of collective software design. The two situations analysed are technical review meetings, and Request For Comments-like procedures in Open Source Software design.

## 1 Introduction

Empirical software engineering (ESE) is concerned with measuring, or estimating, both the effort put into the software (SW) process and the quality of its product. Measurement related to the process reflects the cost (e.g., in person-hours) and is mainly related to the phases as defined in SW process models. Measurement related to the product reflects the SW quality with respect to norms (ISO) or criteria such as reliability, efficiency, usability, maintainability, and portability. A good balance is expected to be found between the effort put into the process and the quality of the final product. However, it is not clear *how* this effort and this quality are related.

We defend the idea that measuring and relating process effort and product quality cannot be performed without a model of cognitive and collective activities involved in SW design, and without measurement of these activities. This is the object of our field, i.e. Cognitive Ergonomics of design. After a brief presentation of its theoretical and methodological foundations, we will discuss a cognitive approach to design activities and its potential to provide new directions in ESE. Then we will present and discuss an illustration of the methodological directions we proposed for the analysis and measurement of cognitive activities in the context of collective software design.

## 2 Cognitive Ergonomics: Theoretical and Methodological Foundations

The main theoretical foundation of cognitive ergonomics [1] is closely related to activity theory. In our approach to design activities, the theoretical framework with respect to the situated and distributed character of cognition is essentially based on information processing psychology and developmental psychology. Our methodological approach is the following. Primary focus is on work in natural settings. We conduct empirical studies, i.e. either field studies, such as observations in the workplace, or ecological laboratory experiments, such as experiments in realistic conditions with real practitioners, realistic tasks, normal constraints, habitual tools, in their usual environments. Data collected consists of, e.g., dialogues, written productions, drawings, and information collected by the designers. This provides relevant data to analyse and evaluate not only the design process, but also its product. Knowledge elicitation techniques and post-hoc interviews based on observational data (e.g. videos, transcripts) may also be used.

## 3 A Cognitive Approach to Design Activities: Potential Directions in ESE

An essential feature of design models developed in cognitive ergonomics is their grounding in empirical evidence: indicators of cognitive design activities gathered in empirical studies are used to specify a model, which, in turn, is repeatedly validated against new empirical data. Existing models concern specific aspects of design (e.g., evaluation, planning and organisation, reuse). In this paper, we will not develop such specific aspects (see e.g. [2],[3]), but present general characteristics of design activities. Indeed, most authors in our field concur in concluding that, whatever the SW process phase, different types of activities are involved in a cyclical —but not systematic— way: problem comprehension, solution generation, solution evaluation, and decision. Furthermore, group management and co-operation activities are also involved all along the SW process.

Empirical studies conducted in other than SW design domains provide evidence of what could be considered "successful" design. Studies in the domain of mechanical design, e.g., have focused particularly on the factors underlying the quality of design ([4],[5]). Characteristics of the activities of the successful designers were the following:

– thorough goal analysis, especially in early design phases;

– initially diverging and then rapidly converging search for solutions, but limiting the amounts of variants, and adopting different perspectives;

– frequent solution evaluation according to comprehensive criteria;

– constant reflection on one's own procedures.

Notice that the designers whose solution quality was evaluated in these studies had all been taught design methodology.

However, empirical studies of design observe that designers often don't proceed in this way. Important drawbacks of design activities are the following (see e.g. [6],[7],[8]):

– limitation in solution search: early choice of a solution without exploration of all alternatives;

– rapid solution evaluation on the basis of just a few criteria: difficulties in taking into account all criteria and their inter-dependencies (constraint management);

– poor involvement of users: requirement identification causes many difficulties; users may be involved in solution evaluation, but they aren't in solution elaboration;

– poor tracing of design rationale: reconstructing the rationale of previous choices causes many difficulties and is time-consuming.

In group management and co-operative activities, there are also inter-comprehension ("cognitive synchronisation") problems and team co-ordination problems, which may affect the cost of the process and SW quality.

A major concern in cognitive ergonomics of design ([2],[9]) is to specify and evaluate methods or tools that would support these activities and overcome their drawbacks. With respect to concerns of measurement in ESE, we propose that such activities be identified and measured as far as they represent the cognitive and collective effort put into the SW process and potentially have an effect on SW quality. For example, breadth of solution search (measured by the number of alternative proposals) and breadth of evaluation (measured by the number and range of criteria used in evaluation, by the number of arguments advanced in favour or against a proposal) could have an effect on the relevance of the chosen solution. We may also assume that if it takes place early in process, with users being involved in this process, less revision of the solution will be required.

## 4  Analysing and Measuring Activities: a Methodological Approach Applied to Collaboration between Designers

Our methodological approach is influenced by our background in Cognitive Ergonomics and Cognitive Psychology and by work in linguistics on dialogues and argumentation. Collaboration is analysed on four levels: from interactions, via sequences and exchanges, to moves. This approach has been used in our previous studies on collaborative design. It can be applied to collaboration between co-designers, working in collocated (e.g. meetings) or distant (e.g., chat), synchronous (e.g., chat) or asynchronous (e.g., discussion lists, email), in oral (e.g., co-presence meetings) or in a textual mode (e.g. chat).

Several steps are distinguished in our methodology. For each step, we will present the principles, the way in which they can be applied, and we will illustrate them by studies in two kinds of situations: (1) face-to-face design situations, in particular technical review meetings (TRMs), and (2) technology-mediated design situations, in particular, Open Source Software (OSS) development.

### 4.1  First Step: Identifying Interactions and Sequences

**Principles**. An "interaction" is a communicative situation characterised by a continuity of actors involved, of the spatio-temporal framework, and of themes discussed. In collocated design, it is typically a meeting, e.g., a co-design meeting. In other contexts, e.g. geographically and temporally distributed design (such as OSS projects), it may correspond to larger temporal situations involving a group of actors. A "sequence" is characterised by the theme under discussion. It is a way of segmenting an interaction into sub-units characterised by one common theme. Typically, a co-design meeting can be decomposed into several thematic sequences. With respect to design, these units could allow measurement of the effort put into the discussion of a theme (generally a sub-problem), whatever the process phase.

**Methods and their limits.** These principles are implemented in methods, which have of course their limits.

*For oral discussions in meetings* (technical review meetings, TRMs). In this situation ([10],[11]), an interaction corresponds to one meeting. Analysing an interaction and its sequences first requires verbatim transcription of the interaction into a textual protocol. Then identifying sequences is done by hand, which generally is difficult because themes under discussion are often implicit. It is also time-consuming. Because of this cost, the method is not adapted to corpus of many, long meetings (e.g., in [10], the method has been applied to seven technical inspection meetings, made up of 148 sequences).

*For textual discussions in chat or discussion lists (OSS projects).* In geographically distributed contexts where discussions take a textual form, there is no need for transcription. Interactions are usually related to a particular task to perform or a document to be commented by the virtual community of designers. For example, we analyse an interaction corresponding to the set of threads related to a Request For Comments procedure in OSS. Sequences are then identified to threads. Finally, the textual sequences can be processed with tools enabling automatic thematic analysis (see, e.g., [12]). Also, sub-themes related to a solution proposal within sequences can be automatically analysed through the "paste as quotation" function that makes explicit the subject and traces the discussion on identical themes.

One may notice that, in contrast with oral discussion meetings, discussions about a theme in OSS projects can spread out over several days, weeks or even years.

### 4.2 Second Step: Identifying Individual Moves

Notice that this second step can be bypassed, as is the case in our study on argumentation in OSS projects.

**Principles.** Individual "moves" are individual contributions to an interaction. A move corresponds to, or is part of, a participant's verbal turn.

A move can be characterised by the type of design activities involved. Design activities highlight the way in which collective design is performed. We have developed a coding scheme on a predicate(argument) basis, details of which may be found in [13]. A verbal move may be an assertion or a request. For example, we distinguish the following predicates.

– Generate: Proposing a new element into the dialogue (a solution, a goal, some inferred data, etc.).

– Evaluate: Judging the value of a subject. This evaluation can either be negative, positive or neutral.

– Inform: Providing new knowledge with respect to the nature of a subject.

– Interpret: Expressing a personal representation of a subject. This representation is expressed through the use of expressions such as "I believe that…", "I think …" or "…maybe…".

The arguments may be Problem data, (Alternative) Solutions, Domain objects, Goals, Domain rules (procedures), Criteria, and Tasks.

**Methods and their limits.** This section presents the methods and their limits.

*For oral discussions in meetings (technical review meetings).* A coding scheme must meet the theoretical criteria of interest and be objective. Objectivity deals with the reliability and validity of the coding scheme, while theoretical interest depends mainly on the domain and on the research goal. Coding categories must be exhaustive and exclusive. A code must be able to model the activities adequately and yet be formal enough to support quantitative analysis. Its definition must be unambiguous in order to different coders attributing the same code to a particular part of the protocol.

Coding is done by hand by an analyst and is time-consuming. In theory, it could be automated, but in reality, this isn't feasible.

*Illustration: Effect of the functional role on effort in evaluation.* In [10], we analysed the effect of the functional role on the relative quantity of individual moves in SW TRMs. Participants in these meetings are all co-designers reviewing SW documents produced by one of them. Their verbal moves may concern different criteria used to assess the documents under study, either form or content criteria. Figure 1 shows that functional role, either project leader (SUP), procedure expert (EXP) or simply co-designer (DEV), has an effect on the moves with respect to the type of criteria. The project leader's moves concern both form and content criteria. The procedure expert's moves mostly concern form criteria, whereas co-designers' moves mostly concern content criteria.

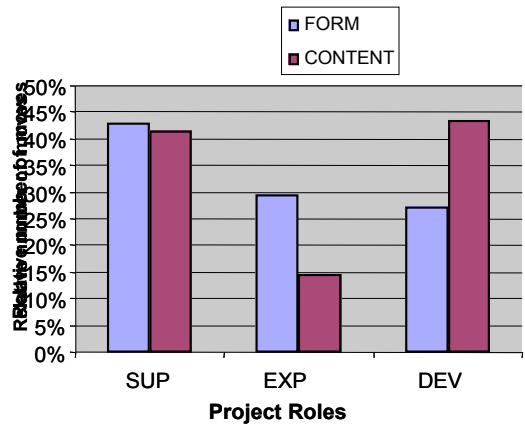

Fig. 1. Effect of functional role on moves relative to form versus content criteria

### 4.3 Third Step: Identifying Exchanges

**Principles.** A verbal "exchange" is a configuration of one or more moves that occurs with a certain frequency in an interaction. It is part of a sequence. We analyse exchanges from two perspectives, as co-operative moves and as argumentation moves. Two examples of co-operative moves are the following.

– Operative synchronisation and co-ordination: This co-operative move fulfils two functions. First, it aims at ensuring that the tasks are shared between the participants involved in the team activity. Second, it aims at ensuring the start, the end, the simultaneity, the sequencing, and the rhythm of the actions to be carried out. Operative synchronisation leads to co-ordination activities.

– Cognitive synchronisation: This move allows the participants in an interaction to ensure that they share two types of knowledge: (i) the same general knowledge about the domain: e.g., technical rules, domain objects, solving procedures; (ii) knowledge about the state of the design: e.g., problem data, state of the solution.

**Methods and their limits.** This section presents the application of our methods and their limits.

*For oral discussions in meetings (technical review meetings).* In our analysis of these meetings, focus has been on co-operative moves (even if there has also been some analysis in terms of argumentation moves). On the basis of the first step coding, each exchange is defined in terms of its different composing activities and their themes.

Patterns (configurations) can be derived using an empirical approach requiring experts in cognitive ergonomics to study the categories of activities characterising the meeting and to derive exchange constituents on the basis of activity patterns. For example, cognitive synchronisation is characterised by Inform or Interpret activities applied on Solutions or Criteria.

We can also validate patterns using a statistical approach, the Lag Sequential Analysis (LSA), which enables to identify categories of activities following one another. This analysis consists in determining whether or not the occurrence frequency of a given category of activities is independent of the occurrence frequency of another category. Sequential structures enable the definition of patterns.

*Illustration: Effort put in cognitive synchronisation, evaluation, and generation in technical review meetings.* In TRMs, we found that one third of the time was spent in cognitive synchronisation —even if the main objective of TRMs is to review (evaluate) a document that represents a state of the SW design project. Furthermore, we found that the designers spend as much time in design (generation) of alternative solutions, which are typical of design meetings, even though, according to the methodology these designers were supposed to use in TRMs, their activities should not include design. This led us to suppose that there is a connection between cognitive-synchronisation and review, on the one hand, and between review and design activities, on the other hand. Further analysis had led to the following conclusions.

There is a bi-directional relationship between review and cognitive-synchronisation activities: (i) a shared representation of the to-be-evaluated object (the object of cognitive synchronisation) is a prerequisite for review activities to occur and (ii) review activities lead to the identification of disagreements concerning the solution under review and/or the evaluation procedure itself: cognitive synchronisation is then triggered by this disagreement that results from review.

The relationship between review and design activities can be explained by the fact that the review of a solution leads the participants to make explicit alternative solutions and to refine the current solution. In the latter case, the participants anticipate activities which, according to their methodology, are supposed to take place in later phases of the global design process.

*For textual discussions in chat or discussion lists (OSS projects).* Second step coding has been bypassed. In this study, we are primarily focussed on argumentative exchanges. We currently use a manual annotation of the corpus to distinguish between proposals, arguments (pro or cons) and decisions. We also search manually for significant (and easily usable) linguistic markers denoting these different activities. However, we plan to adapt existing tools (annotation, Natural Language processing) to partially automate the process. In the future, such tools could be used during the design process to inform and assist the designers in managing the design process.

## 5   Discussion

Besides measuring concrete features of the design outputs and standards, assessing SW design effort related to quality should also reflect the "cognitive effort" involved in the development process. This paper has proposed and illustrated a method for the measurement of some aspects of this "cognitive effort". Such an approach gives rise to at least two potential directions for SW empirical evidence measurement. First, it enables to quantify the cognitive activities associated with each project development, in order to complement other, concrete and output-related indicators when attempting to empirically compare various developments. Second, it enables to investigate the influence of these activities on the quality of the process and the final product.

An application of our method not presented in this paper consists in measuring the breadth of solutions that have been envisioned in solution development, and of criteria that have been applied in evaluating design proposals. Given that one knows that solution breadth, especially at the start of a design project, is a predictor of quality, and that one may suppose that the breadth of evaluation also constitutes such a predictor, these measures may be useful.

Different types of analysis presented in this paper allow to assess the proportion of a design meeting duration that is allocated to a particular activity (cognitive synchronisation), or a particular object (type of evaluation criteria implemented according to one's functional role). The transition from this type of results to an assessment of quality, or even effort, is not immediate. The time devoted to an activity A is not the only predictor of the quality of the result with respect to A (a great proportion of a meetings' duration having been spent in solution generation doesn't forecast the quantity, let alone the quality of solutions having been generated). Our analysis shows that cognitive synchronisation activities are a prerequisite of evaluation. According to the distribution of cognitive synchronisation over a meeting, however, the same total proportion of cognitive synchronisation distributed differently may lead to different results, that is, to SW products of different quality.

Combining classical ESE methods with those presented in this paper is to be investigated in order to establish a link between, on the one hand, cognitive activities and design objects (solutions, criteria) and, on the other hand, their impact on effort and quality of SW design projects —in order to establish, ultimately, the optimum trade-off between (cognitive) effort and the quality of the design product.